# Does the universe obey the energy conservation law by a constant mass or an increasing mass with radius during its evolution?

**Akinbo Ojo**
Standard Science Centre
*P.O. Box 3501, Surulere, Lagos, Nigeria*
taojo@hotmail.com

**Abstract**
How the energy conservation law is obeyed by the universe during its evolution is an important but not yet unanimously resolved question. Does the universe have a constant mass during its evolution or has its mass been increasing with its radius? Here, we evaluate the two contending propositions within the context of the Friedmann equations and the standard big bang theory. We find that though both propositions appeal to the Friedmann equations for validity, an increasing mass with increasing radius is more in harmony with the thermal history of the big bang model. In addition, temperature and flatness problems that plague the constant mass proposal are mitigated by the increasing mass with radius proposal. We conclude that the universe has been increasing in mass and radius in obedience to the energy conservation law.

**Key words**: cosmology, big bang, energy conservation, temperature problem, flatness problem

## I. Introduction

Since the formulation of the standard big bang model [1,2], the question of how to proceed with the universe's evolution without seriously violating the energy conservation principle has been an early area of contention and till date it is still not a unanimously settled issue. While in one view it is speculated that the universe has been evolving with a constant mass but increasing radius, (e.g. Lemaitre [3]), on the alternative view it is variously mentioned in passing that total energy being the sum of the positive energy of matter and the negative energy of the scale factor or radius is what is conserved as the universe evolves [4-6]. The implication of the first proposition is that the matter-energy content of the universe is a conserved quantity, starting from a massive object of zero radius and thus infinite density to future eras with increasingly reduced energy density but constant mass. In the event of a time-reverse, collapse will equally be to an object of zero radius which still retains its matter-energy content and will thus be of infinite density again. The second proposition implies that if the total energy sums to zero, when the radius is zero, the matter-energy content too will be zero. This proposition is incompatible with theories of a universe starting from 'nothing', (e.g. [7]). During the universe's evolution, the second proposition also implies that if there is an increasing radius with negative energy, the positive matter-energy content would have to be increasing as well to keep the total energy sum zero. In spite of the two propositions deriving their basis from interpretations of the Friedmann equations [8], they are mutually exclusive. Only one can be correct. The relevant Friedmann equations are

$$3R'^2/R^2 + 3kc^2/R^2 - c^2\Lambda = 8\pi G\rho \qquad (1)$$

$$-2R''/R - R'^2/R^2 - kc^2/R^2 + c^2\Lambda = 8\pi Gp/c^2 \qquad (2)$$

where R is the scale factor, the accent, ( $'$ ) denotes differentiation with respect to time, k is the curvature parameter, c is light velocity, $\Lambda$ is the cosmological constant, G is the gravitational constant, $\rho$ is the matter density and p is the pressure. The equations themselves derive largely from the field equations of General relativity (GR).

In the following sections, we first recap the basis for the constant mass scenario within the context of the Friedmann equations, then in the next section we look at the same equations for information about increasing mass with radius. After this we revisit the big bang model and the derivation of its thermal history in section IV. Having done the foregoing, we are able to discuss the two propositions and their consistency or lack of it with the big bang theory in section V. Concluding remarks are made in the last section.

## II. Constant mass with increasing radius.

Probably the first to publish a peer-reviewed paper seeking to address the associated energy conservation concerns in the newly proposed big bang theory was Lemaitre. His paper ended with the conclusion that the universe was evolving with a constant mass but increasing radius [3]. Others have made similar inferences drawing strength from the Friedmann equations, particularly Eq.(2) which unlike the first contains a pressure term, p.

In brief, by differentiating the first Friedmann equation with respect to time, then combining it with the second and finally multiplying by $R^3$ arrives approximately at

$$\partial(R^3\rho)/\partial t + p\partial(R^3)/\partial t = 0 \qquad (3)$$

The constancy of the total energy in a closed system is stated by the first law of thermodynamics and is usually written

$$\partial U + p\partial V = \partial Q \qquad (4)$$

where $\partial Q$ is the heat transferred into or out of the system, $\partial U$ is the change in internal energy, p is pressure and $\partial V$ is the change in volume. For adiabatic changes which must apply to the universe since no heat can be transferred into or removed from the system, $\partial Q = 0$. Therefore, Eq.(4) will be

$$\partial U + p\partial V = 0 \qquad (5)$$

This equation is strikingly similar to Eq.(3), justifying the proposition of Eq.(3) as the energy conservation law applicable during the universe's evolution.

Equation (3) implies that if the expanding universe can have a pressure, then when p>0, the total energy per comoving volume decreases with time and when p<0, the total energy per comoving volume increases with time. The possibility of a false vacuum state where p<0 and negative formed the basis for early inflationary proposals [9,10] to modify the standard big bang model, with an enormous increase in matter-energy content accompanying an exponential increase in volume. For a pressureless universe where p = 0, when volume increases, Eq.(3) shows that $\partial(R^3\rho)/\partial t$ will be zero and therefore there will be no change in matter-energy content with time. This is the basis for the constant mass with increasing radius proposal. This constancy of the total mass within the comoving volume is also sometimes discussed as the 'continuity equation'. That is

$\rho R^3$ = constant  (6)

Thus if the pressure has been negligible or zero, at least after any possible inflationary epoch, then it is implied that the matter-energy content of the observable universe must have been conserved right from the beginning or soon after an inflationary scenario. The current estimate for this mass is ~ $10^{52}$kg (i.e. $10^{69}$J).

At this point it may be pertinent to note for consistency arguments that even though the mass of the system has become interchangeably used with its internal energy in arriving at the above conclusions, in the original statement of the energy conservation law, Eq.(4), which Eq.(3) simulates, reference is actually to the energy of motion of the constituent particles which depends on their temperature, i.e. the 'internal energy'. Reference is not to the mass possessed by the particles. To illustrate this consistency argument, when a given mass of gas is adiabatically frozen to absolute zero, although its internal energy becomes zero (neglecting any zero-point energy), the mass of the gas does not become zero. Some caution may therefore need to be exercised in interchangeably equating mass with internal energy in energy conservation laws as intended by Eq.(3). Having noted this precaution, if by argument from Einstein's E = mc$^2$, change in internal energy can be equated to change in mass, then for a pressureless universe, Eq.(3) shows that the universe's mass will be conserved and would not change during its adiabatic evolution.

With mass being a constant, matter-energy density, $\rho$ will vary as the inverse cube of the radius, i.e. as r$^{-3}$. However this would only be in the matter-dominated era, as in the radiation-dominated era, it is suggested that because of the cosmic red-shift resulting from expansion, the frequency and energy per photon will scale as r$^{-1}$, matter-energy density, $\rho$ would therefore vary as r$^{-4}$ in that era. This is the standard view which we refrain from criticizing here.

From black body radiation laws which we discuss in more detail later, energy density when in the form of radiation is related to the fourth power of the absolute temperature, T, i.e. $\rho = a$T$^4$, where *a* is the radiation constant (see Eq.(17) later. Thus, in the radiation-dominated era, if $\rho$ varies as r$^{-4}$, the temperature, T will be expected to vary as the inverse of the radius of expansion, i.e. T varies as r$^{-1}$ during that era.

### III. Increasing mass with radius

The Friedmann equations, especially the first can also be used to describe an increasing mass with radius. If as is common, we make the following assumptions that $\Lambda$ is negligible, R′/R approximates to the Hubble parameter, H and that for a universe beginning from nothing its total net energy is zero and it will therefore not be closed or open, making the curvature parameter k equal to zero, then from Eq.(1) we can arrive at

$$1/H = \sqrt{(3/8\pi G\rho)} \tag{7}$$

The inverse of the Hubble parameter is the expansion time, t and the distance light can cover during the expansion time is the radius of the observable universe, r, i.e. r = ct. Thus, $1/H = t = r/c$. We can from Eq.(7) therefore write a relationship between matter density and expansion time as

$$\rho = 3/8\pi Gt^2 \tag{8}$$

and also a relationship between matter density and the radius of the visible universe as

$$\rho = 3c^2/8\pi Gr^2 \tag{9}$$

The relationships in Eqs.(8) and (9) show that matter density reduces with expansion time and increase in the radius.

Equation (9) shows that as r increases with time, $\rho$ reduces as $r^{-2}$. If mass is constant, density varies and reduces as $r^{-3}$. Thus inherent in the relationship in Eq.(9) is an implied suggestion that the matter-energy content may not be constant and the mass of the universe may be increasing linearly with increasing radius.

To attempt to give a quantitative value to this mass increase with radius, we can replace the matter density, $\rho$ in the first Friedmann equation with the mass, M in the observable universe of volume, $4\pi r^3/3$, where r is the radius of the observable universe. If R′/R is the Hubble parameter and its inverse is the expansion time, t, then we have

$$r^3/2G(1/t^2 + kc^2/R^2 - c^2\Lambda/3) = M \tag{10}$$

With a negligible Λ, the assumption of flatness, k = 0 and r/t being the light velocity, c, solving Eq.(10) eventually gives us

$$rc^2/2G = M \qquad (11)$$

Since the radius of the observable universe is not constant but increases with time and G and c are constants with values $6.67 \times 10^{-11}$ $kg^{-1}m^3s^{-2}$ and $3 \times 10^8 ms^{-1}$ respectively, if Eq.(11) holds for a universe with the stated assumptions of negligible Λ and k = 0, then for every metre change in the radius of the observable universe, there must be a $6.75 \times 10^{26}$ kg change in mass. We can also express this in joules, if $E = mc^2$, a $6.07 \times 10^{43}$ J change in matter-energy content must occur for every metre change in radius. Therefore, for a universe obeying the energy conservation law by increasing mass with radius, from Eq.(11) at Planck radius ~ $10^{-35}$ m, its mass will be about ~ $10^{-8}$ kg and at radius ~ $10^{26}$ m, its mass would be about ~ $10^{52}$ kg.

For comparison with the constant mass proposal, the relationship between temperature and the radius during the radiation-dominated era is noteworthy. Since $\rho$ is proportional to $aT^4$, then putting this in Eq.(9), temperature, T will vary as $r^{-½}$ for the increasing mass with radius proposal.

If both M and r increase simultaneously and yet for a closed system there is no change in the total energy which remains conserved, then in Eq.(11) negativity must be attributed to one of the two variable terms and positivity to the other. By convention positivity is usually ascribed to matter-energy content and negativity to the radius. The energy conservation law can therefore be expressed by stating that the total energy of the universe is the sum of the positive energy of matter and the negative energy of the radius and it is a conserved quantity. For a universe starting from nothing, total energy will be conserved at zero at all epochs and any change in radius must always be accompanied by a commensurate change in mass.

There are however some aspects worth examining for consistency purposes. Eqs.(8) and (9) seem to suggest that at time zero and when the universe was of zero radius, density would be infinite. This conflicts with the proposal here that positive energy of matter and the negative energy of the radius always sum to zero. A further analysis of the Friedmann equations as was done above shows that using $M/(4\pi r^3/3)$ instead of $\rho$ in Eqs.(1),(8) and (9), we obtain $M = r^3/2G(R'^2/R^2 + kc^2/R^2 - c^2\Lambda/3)$, $M = r^3/2Gt^2$ and $M = rc^2/2G$ respectively. From this perspective, M will be zero when r is zero. Since r is zero when t is zero, the beginning would therefore be equivalent to a nothing state and not an infinitely dense one. In this sense, the Friedmann equations can apply up to time zero and still be compatible with the proposal in this section.

A second point worth mentioning here is the psychological difficulty understanding where additional mass could be coming from as radius increases. Some kind of clarifications therefore seem needed whether the law of conservation of mass is a local or a cosmological law, especially in the light of cosmological theories of creation from nothing, since a universe created from nothing cannot obey this law.

## IV. A brief overview of the thermal history of the big bang model

Having looked at the theoretical background of the two contending proposals how energy conservation law is obeyed, we now briefly revisit the big bang theory which both support. In particular, we examine relevant aspects of the thermal history of the big bang model for information about its matter-energy content.

The big bang theory proposes the universe having a very high matter-energy density in the past which has been reducing due to expansion. As a result, temperature was also very high in the past and has since been reducing. Lemaitre, one of the proponents of the constant mass view is recognized as one of the architects of the theory. The major mathematical framework for the theory is based on the Friedmann equations. The other relevant aspect of physics required to understand the thermal characteristics, especially during the radiation-dominated era are aspects of the laws of black body radiation. We briefly look at these.

The most relevant Friedmann equation remains Eq.(1). Given that the area of focus here is the thermal history and the inverse of H is the expansion time, t, we can write this Friedmann equation as

$$3/8\pi G \ (1/t^2 + kc^2/R^2 - c^2\Lambda/3) = \rho \tag{12}$$

If $E = mc^2$, then matter density, $\rho$ expressed in joules to give energy density, $E_D$ will make Eq.(12)

$$3c^2/8\pi G \ (1/t^2 + kc^2/R^2 - c^2\Lambda/3) = E_D \tag{13}$$

Equations (12) and (13) vividly show that at the earliest expansion times, matter-energy density must have been very high and this reduces as the period of expansion increases.

Now if the universe is 'all there is', the universe's radiation could not have originated from another radiating body and transmitted through it or reflected from it. The radiation would therefore be intrinsic to it, possessing the characteristics of a black body and obeying the laws associated with such radiation [11,12]. Among such laws is the Stefan-Boltzmann law

$$P/A = \sigma T^4 \qquad (14)$$

where P is the power or the energy radiated per second, A is the surface area, $\sigma$ is Stefan's constant, (5.7 X $10^{-8}$ $Wm^{-2}K^{-4}$) and T is temperature in Kelvin.

The energy density, $E_D$ for radiation of all wavelengths in a given volume, can be related to the Stefan-Boltzmann law by

$$\sigma T^4 = c/4 \text{ X } E_D \qquad (15)$$

where c is light velocity. Rearranging, we can write

$$E_D = 4\sigma/c \text{ X } T^4 \qquad (16)$$

Since the expression $4\sigma/c$ gives a constant, we have

$$E_D = aT^4 \qquad (17)$$

where *a* has the value 7.56 X $10^{-16}$ $Jm^{-3}K^{-4}$ and is known as the radiation constant. Thus the amount of energy in a given volume of space, when in the form of radiation can be related to temperature.

Replacing $E_D$ with $aT^4$, Eq.(13) becomes

$$3c^2/8\pi G \, (1/t^2 + kc^2/R^2 - c^2\Lambda/3) = aT^4 \qquad (18)$$

or by a further arrangement

$$T^4 = 3c^2/8\pi Ga(1/t^2 + kc^2/R^2 - c^2\Lambda/3) \qquad (19)$$

Equation (19) is the equation for the thermal history of the universe during the radiation-dominated era whether flat, open or closed. If we assume a negligible $\Lambda$ and that the curvature term, $kc^2/R^2$ is zero or negligibly positive or negative in the early universe, then taking the square root of Eq.(19) gives us

$$T^2 = \sqrt{(3c^2/8\pi Ga)} \times 1/t \qquad (20)$$

The values in bracket are constants, so we can write

$$T^2 = (4.6 \times 10^{20}) \times 1/t \qquad (21)$$

This equation allows us to quickly determine the thermal character of given epochs of the early radiation-dominated universe. The above equations are generally accepted in cosmology and black body radiation physics.

We now select three sample epochs after the beginning of expansion (ABT) for further study. The Planck epoch $\sim 5.4 \times 10^{-44}$ seconds ABT, $10^{-32}$ seconds ABT which will be after any possible inflation and $10^{-10}$ seconds ABT.

From Eq.(21), the temperatures according to the big bang model at these sample epochs are approximately $10^{32}$K, $10^{26}$K and $10^{15}$K respectively. Under our discussion section, we later examine the corresponding energy densities which can be derived from Eq.(17) or calculated directly from Eq.(13) for further information about the implied matter-energy content at these epochs.

### V. Discussion

To put our cosmology on a sounder footing, it is an important task to determine which of the two propositions of how to preserve the law of energy conservation is the more consistent with theory and observation. This may not be so straight forward as they share some similarities. They both predict a state of high energy density and thus high temperature in the past with temperature falling in the future as the universe expands. They also both appeal to the Friedmann equations for validity. Despite the similarities, the two propositions cannot both be correct.

To assist us in evaluating their consistencies, we first look at the theoretical scenarios for the beginning, even though these are not directly testable. For a universe that has arisen from an object of infinite density and zero radius, the second proposition would be ruled out since total energy of matter and radius would then not sum to zero at the beginning. On the other hand, a universe created from nothing and collapsing to nothing would seem to rule out a constant mass proposal but would favour the proposal that the energy of matter and the negative energy of the radius always sum to zero and the matter-energy content increases as the universe increases in its radius.

The standard big bang model does not differentiate for us what exactly the situation at time zero is. Do we have 'nothing', i.e. no radius, no matter-energy and thus absolute zero temperature or do we have zero radius but positive matter-energy content and thus infinite temperature? As earlier mentioned, at first glance, the big bang model through Eq.(9) seems to tell us that at zero radius, density was infinite. However, if we use mass per volume in place of density as in Eq.(10), the big bang model seems to tell us that when radius is zero, mass would also be zero. Despite this equivocation, the model can guide us what happens afterwards from the Planck epoch and we have a fairly reliable and accepted thermal history for the radiation-dominated era.

From Eqs.(19-21), we can find a relationship between temperature and the increasing radius of the observable universe. Since r is the distance light can cover during the expansion time, if we replace t with r/c we can deduce that the temperature, T in the radiation-dominated era varies as $r^{-½}$. The significance of this is that the thermal history of the big bang model, at least during the radiation-dominated era seems biased towards the thermal history earlier identified with a universe whose mass is increasing with its radius. As can be recalled, a constant mass universe would have its temperature varying as $r^{-1}$ during the radiation dominated era.

By far the most likely guide to choose between the two proposals appears to be the presence of associated paradoxes. Here, we discuss two. A less discussed temperature problem and the common flatness problem, first highlighted by Dicke and Peebles [13]. The two problems are related being both associated with the matter-energy content and energy density of the universe.

A temperature puzzle can be formulated which seems to rule out that mass was anywhere near the observed value now $\sim 10^{52}$ kg (i.e. $10^{69}$ J) during the radiation-dominated era up to $10^{-10}$ seconds ABT. To illustrate this, from Eq.(17) knowing the radius of the observable universe at the Planck epoch $\sim 10^{-35}$ m, if the matter-energy content was $10^{69}$ J, which would be present in the form of radiation since matter is unstable at temperatures above $10^{15}$K, the energy density at that epoch would be above $10^{174}$Jm$^{-3}$ translating to a temperature $10^{47}$K and not the modelled ~$10^{32}$ K.

There are various versions of the inflation scenario. In most versions, our observable universe is just one of a large number of spherules within an exponentially attained 0.1m radius. In some scenarios, as many as $10^{67}$ spherules may be contained in this inflated sphere of 0.1m radius. If our spherule has causally connected regions based on the observed homogeneity in our universe, this will be of radius ~ $10^{-24}$m, being the maximum distance light could have travelled since the beginning ~ $10^{-32}$s ABT. Energy density would have been ~ $10^{141}$Jm$^{-3}$, if matter-energy content after inflation is present and conserved as ~$10^{69}$J, translating to a temperature above $10^{39}$K and not the modelled ~ $10^{27}$K.

The physics described at $10^{-10}$s ABT is perhaps less speculative. The universe then will have a radius ~ 0.1m. If the mass of the universe now was also the same as it was then, the energy density will be $10^{72}$Jm$^{-3}$, translating to a temperature ~ $10^{22}$K and ambient energy ~$10^9$GeV at that epoch, far higher than the $10^{15}$K and an ambient energy ~ $10^2$GeV which would be the ideal condition for the Salam-Weinberg phase transitions separating the electroweak into electromagnetic and weak nuclear forces as predicted by the big bang model. These temperature problems would therefore seem to rule out the universe having the mass it currently possesses during those earlier radiation-dominated eras.

From the model temperatures of the big bang model, we can estimate what the consistent matter-energy contents would be at the three sample epochs. For an early universe at the Planck radius $\sim 10^{-35}$m, the standard model predicts a model temperature $\sim 10^{32}$K. From black body radiation laws, the corresponding energy density is $\sim 10^{112}$Jm$^{-3}$. Knowing the volume of the universe at that epoch, the matter-energy content required for accurately modelling the standard big bang at that epoch would be less than $\sim 10^{10}$J. Any matter-energy content above this would make the universe much hotter than $10^{32}$K!

A temperature $\sim 10^{27}$K is predicted after inflation. From the radiation density formula, the matter-energy content that will be compatible with this predicted state of the observable universe with radius about $10^{-24}$m will be $\sim 10^{21}$J.

At $10^{-10}$s ABT, the standard model predicts a temperature $10^{15}$K when the observable universe had a radius $\sim 0.1$m. This translates to an energy density about $10^{44}$Jm$^{-3}$. The matter-energy content required for a universe of this size to have consistency with the standard big bang model temperature will be about $10^{42}$J. Notable from all the above is that from the Planck epoch to $10^{-10}$s ABT, the matter-energy content consistent with the model temperatures of the observable universe according to the big bang theory seems to have been increasing. Calculating with less approximation suggests $\sim 10^{43}$J per metre change in radius occurring during the three consecutive sample epochs. This thermal feature of the big bang thus appears to support the increasing mass with radius proposal. To summarize the temperature problem, a mass of $10^{52}$ kg would make the curvature term, $kc^2/R^2$ completely dominant in Eq.(19) and the thermal history in the early eras would be completely different from that modelled by the big bang theory.

The second puzzle is the common flatness problem. The essence of this puzzle concerns the parameter, $\Omega$, usually expressed as the ratio of actual matter density, $\rho$ to the critical density, $\rho_c$. For clarity of discussion, since the volumes under reference are the same, $\Omega$ is also the ratio of the actual matter-energy content in the universe, M to the matter-energy content required to close the universe, $M_c$. It can be shown that $M_c$ is the same as the mass required for the system to be within its gravitational or Schwarzschild radius. Current estimates place $\Omega \sim 0.04$, excluding dark matter and $\Omega \sim 0.3$ inclusive of dark matter.

Despite the uncertainty and seemingly wide range in the value of $\Omega$, Dicke and Peebles have pointed out that for $\Omega$ to even be within this vicinity of one today, it must have been even nearer the vicinity of one very early in the universe's evolution. The basis given for their assertion was that were this not so, the universe would have either been dramatically open now after $\sim 10^{60}$ Planck times of expansion or it would have closed already. In other words, since the universe is yet to be either significantly closed or open based on the observed closeness between the mass of the universe now and the mass required to close it, this approximation must have held and would even have been more in earlier eras since any differences of $\Omega$ from one get magnified by at least one decimal place with each Planck time that elapses. Here in this paper, we are concerned mainly with what this riddle about the cosmic history of $\Omega$ implies for the constant mass and the increasing mass with radius proposals for energy conservation and not the exact value of Ω today to which most current efforts are directed.

The critical density, $\rho_c$ required to keep the universe closed at any epoch and at which k = 0 can be defined from Eq.(1) as $3H^2/8\pi G$, for a negligible Λ. For the purpose of clarity, since H = 1/t = c/r, we also write the critical density in terms of expansion time, t as

$$\rho_c = 3/8\pi Gt^2 \tag{22}$$

and in terms of the radius of the observable universe, r as

$$\rho_c = 3c^2/8\pi Gr^2 \tag{23}$$

Using the critical mass, $M_c$ contained within the observable volume, $4\pi r^3/3$ in place of $\rho_c$ in Eq.(23), we have

$$M_c = rc^2/2G \tag{24}$$

It can be seen therefore that the critical mass required to provide the critical density is the same as the mass required for the system to be within the boundary of its Schwarzschild radius.

We can thus use either $\rho_c = 3c^2/8\pi Gr^2$ or $M_c = rc^2/2G$ to determine the critical density limit or the critical mass at different epochs for respective universes in a simple way, once the radius is known.

The parameter, Ω, i.e. $\rho/\rho_c$ or $M/M_c$, determines whether the universe is open, flat or closed. An Ω value above, equal to or below one implying a closed, flat or open universe respectively. When the mass is more than that required to keep the system within its Schwarzschild radius, i.e. when $M > rc^2/2G$, $\Omega > 1$. When the mass is below the critical mass and is not enough to keep the body within its Schwarzschild radius, i.e. when $M < rc^2/2G$, $\Omega < 1$. And when the mass is exactly equal to that required to keep the body within its Schwarzschild radius, i.e. $M = rc^2/2G$, $\Omega = 1$ exactly. Similarly, defining $\Omega$ with density, when $\rho > 3c^2/8\pi Gr^2$, i.e. $\rho > \rho_c$, $\Omega > 1$, when $\rho = 3c^2/8\pi Gr^2$, i.e. $\rho = \rho_c$, $\Omega = 1$ and when $\rho < 3c^2/8\pi Gr^2$, i.e. $\rho < \rho_c$, $\Omega < 1$.

We can now discuss the relevance of the flatness problem to our topic with more clarity. From $M_c = rc^2/2G$, the critical mass that makes $\Omega = 1$ at the Planck epoch with radius $\sim 1.62 \times 10^{-35}$m is $\sim 10^{-8}$ kg. Today at about 10 billion years with radius $\sim 10^{26}$m, the critical mass now from $M_c = rc^2/2G$ to make $\Omega = 1$ is $\sim 10^{52}$kg. Knowing the respective volumes, we can calculate what the corresponding critical densities will be. We can also get the same results using $\rho_c = 3c^2/8\pi Gr^2$ or $\rho_c = 3/8\pi Gt^2$. At the Planck epoch, the critical density, $\rho_c$ is about $6.14 \times 10^{95}$kgm$^{-3}$. Today, at expansion time about 10 billion years, $\rho_c \sim 1.8 \times 10^{-26}$kgm$^{-3}$ (or $1.8 \times 10^{-29}$gcm$^{-3}$ in other commonly used units). Knowing the respective volumes, we can also calculate what the corresponding critical masses will be. A consistency with the current values for critical density estimated using $3H^2/8\pi G$ and the observed estimate for the Hubble constant is demonstrated.

The mass of the universe that has been deduced from observation is within a magnitude of an order or two of $10^{52}$kg. If the universe had this mass all along and it is a constant, then knowing the critical density and critical mass at the Planck era as stated above, we see that $\rho \gg \rho_c$ and $M \gg M_c$, i.e. $M \gg rc^2/2G$ making $\Omega \gg 1$ at that era and the universe would have closed much earlier on. This dilemma was part of the motivation for inflationary scenarios so that the radius of the universe is exponentially increased to bring the state near M ~ $rc^2/2G$, i.e. $\Omega$ ~ 1, in order to prevent the early closure for an observable universe with that amount of mass.

During the inflationary scenario, the universe as a whole has its radius increased from some magnitude above the Planck radius to 0.1m or $10^{10000000000}$m depending on the version. Our observable universe becomes just a tiny speck of the whole. After inflation, about $10^{-32}$ seconds ABT, our observable homogenous universe would come to have a causally connected region of radius ~ $10^{-24}$m. From $\rho_c = 3c^2/8\pi Gr^2$ and $M_c = rc^2/2G$, we can determine what the critical density or critical mass would be for the resulting observable universe immediately after inflation. This comes to $\rho_c$ = 1.61 X $10^{74}$kgm$^{-3}$ or $M_c$ = 675kg. Thus while inflation is a device proposed to remedy the difficulties associated with an enormously massive beginning and bring the system from $\Omega \gg 1$ to $\Omega$ ~ 1, in the final analysis inflation will fail to fully achieve this if the mass of the observable universe is anywhere close to $10^{52}$kg after inflation since $\Omega$ (i.e. $\rho/\rho_c$ or $M/M_c$) would still be above $10^{49}$. To obtain $\Omega$ ~ 1 immediately after inflation, the mass of the observable universe after the event must not be more than 675kg!

A dilemma therefore seems to exist for inflation theory with both the constant mass and the increasing mass with radius proposal. With the constant mass proposal, which inflation was designed to rescue, we see above that inflation fails to achieve flatness if the mass of the observable universe is $10^{52}$kg after the event. With the increasing mass with radius proposal, if the universe at the Planck time had a mass ~ $10^{-8}$ kg and a density 6.14 X $10^{95}$ kgm$^{-3}$, then these being the critical mass and critical density for a universe of that size, the universe was therefore already flat at that epoch and an inflationary event that causes flatness would become unnecessary and superfluous.

Leaving the issue of the difference between M and $M_c$ in order to determine the exact value of $\Omega$, what Dicke and Peebles have shown and stated in the flatness problem is that for the actual mass of the universe to approximate the critical mass now, then the two must have been even further approximate to each other in earlier eras, i.e. $\rho \sim \rho_c$ or M ~ $M_c$ at each past epoch, to make the cosmic history of $\Omega$ to be within the vicinity of one till the present time. Since the critical mass that gives the critical density increases with radius, it follows that the actual contained mass must also have been increasing to make this approximation between M and $M_c$ agree with the flatness riddle and the current observational measurements of $\Omega \sim 1$.

At the Planck era, the critical mass as shown above is $10^{-8}$ kg. At radius ~ $10^{-24}$m and 0.1m, the critical masses of the observable universe will be 675kg and 6.75 X$10^{25}$kg respectively. The flatness puzzle therefore implies that for $\Omega$ to have always been within the vicinity of one, the actual mass, M of the observable universe has in a wondrous way always virtually approximated the critical mass, $M_c$ at each past epoch and thus must have been increasing by approximately 6.75 X $10^{26}$kg per metre change in radius. The wonder in the riddle is removed by the energy conservation principle if the mass increase with radius is in obedience to it.

## VI. Concluding remarks

The knowledge that the mass of the universe now seems to be a mixture of radiation, matter and dark matter forms has become fairly established. But the exact way the universe obeys the energy conservation law remains fundamental to our understanding of cosmic evolution. Has all that mass always been there, the proportion of the various forms depending on the prevailing temperature that permits their stability? Or does additional mass appear with increasing radius, taking the form dictated by the energy densities and ambient temperatures prevailing at its appearance?

The psychological barrier, where additional mass can come from, especially when considered alongside the law of conservation of mass, must undoubtedly have played a significant role in the speculations leading to the constant mass hypothesis. Alternative ideas of creation from nothing which later emerged have however become increasingly popular. Despite remaining hesitations about where additional mass could be coming from, on the overall balance of consistency, we conclude that the energy conservation law is better obeyed by means of increasing mass of the universe with its radius. The bases for our conclusion include compatibility with the Friedmann equations with fewer assumptions and improvisation, better harmony with the thermal features of the standard big bang model and the mitigation of the temperature and flatness problems which characterize the constant mass proposal.

The findings here are obtained almost entirely within the context of General relativity, which has a wide acceptance in the scientific community. However, not too different conclusions can be described outside GR, as for example based on the behaviour of a primeval photon fluctuating from nothing [14].

**Acknowledgements**
I thank Alex Vilenkin for ideas shared with me in personal correspondence.